# Cognitive Aging and Labor Share

B.N. Kausik[1]

Aug 24 2023

## Abstract

Labor share, the fraction of economic output accrued as wages, is inexplicably declining in industrialized countries. Whilst numerous prior works attempt to explain the decline via economic factors, our novel approach links the decline to biological factors. Specifically, we propose a theoretical macroeconomic model where labor share reflects a dynamic equilibrium between the workforce automating existing outputs, and consumers demanding new output variants that require human labor. Industrialization leads to an aging population, and while cognitive performance is stable in the working years it drops sharply thereafter. Consequently, the declining cognitive performance of aging consumers reduces the demand for new output variants, leading to a decline in labor share. Our model expresses labor share as an algebraic function of median age, and is validated with surprising accuracy on historical data across industrialized economies via non-linear stochastic regression.

JEL D63, E22, E23, E24, J24, O33, O41

[1] Conflict disclosure: unaffiliated independent.
Author's bio: https://sites.google.com/view/bnkausik/  Contact: bnkausik@gmail.com
Thanks to D. Autor, F. Baskett, P. Bockerman, J. Jawahar, T. Johnson, R. Krishnan, D. Raval  A. Salomons, P. Tadepalli & K. Suresh

# Introduction

Labor share, the fraction of economic output accrued as wages, is inexplicably declining in industrialized countries, raising the question of whether automation will displace human labor. Prior work, such as Frey and Osborne (2013), Korinek and Stiglitz (2020), Brynjolfsson (2022) and Acemoglu (2022) are pessimistic, while others such as Arntz, Gregory and Zierahn (2017), Bessen (2019), Autor et al (2020a), and Basu (2022) offer reasons for optimism. Nevertheless, there is sustained evidence that labor's share of economic output is declining in industrialized countries. Fig. 1 shows US labor share per the series "Shares of gross domestic income: Compensation of employees, paid: Wage and salary accruals: Disbursements: to persons (W270RE1A156NBEA)" from the St. Louis Federal Reserve.

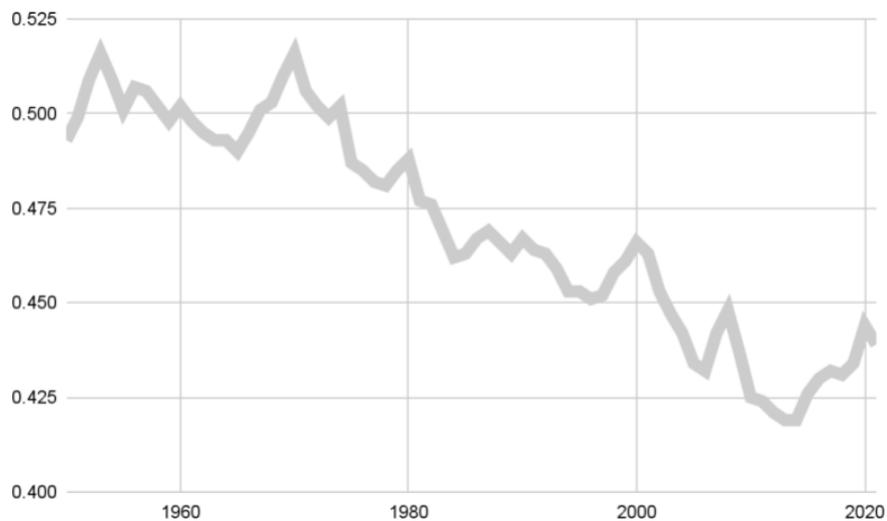

Fig. 1. US labor share 1948-2021 (data source: St. Louis Fed)

Fig. 1 suggests that US labor share was stable between 1948 and 1970 but declined since then. A similar decline is observed in other industrialized countries.

The cause of the decline is the subject of thousands of papers as surveyed by Grossman and Oberfield (2021), ranging across (1) technological change, e.g., Karabarbounis and Neiman (2014), Oberfield and Raval (2014); (2) globalization, e.g., Elsby et al. (2013), Autor et al. (2020b); (3) concentration of market power, e.g., Barkai (2020), Autor et al. (2020b); (4) decline in unionization, e.g, Bentolila and Saint Paul (2003), Benmelech et al. (2020); (5) demographic changes in the workforce, e.g., Glover and Short (2020), Hopenhayn et al. (2018); with the conclusion that an explanation for the decline in labor share remains elusive.

A popular perspective is that human labor and automation compete for their respective share of economic activity, e.g. Acemoglu and Restrepo (2018), Acemoglu and Restrepo (2019), Autor et al (2022), and that automation inexorably wins, driving down labor's share of economic output. In contrast, our perspective is that labor share is determined via a competition between human consumers[2] and human suppliers, and is influenced by biological factors related to the median age of the population, which is increasing in industrialized countries as shown in Figs. 2a and 2b (data source: https://ourworldindata.org/).

---

[2] We use the term "consumers" to refer to all participants in the economy.

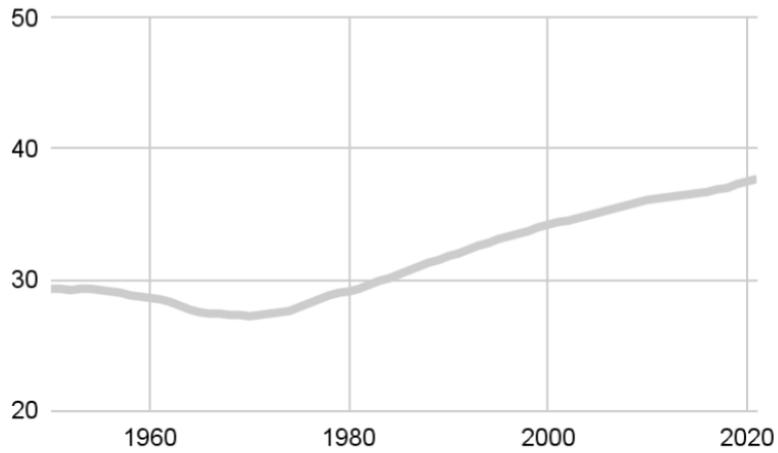

Fig. 2a. US median age 1950-2021

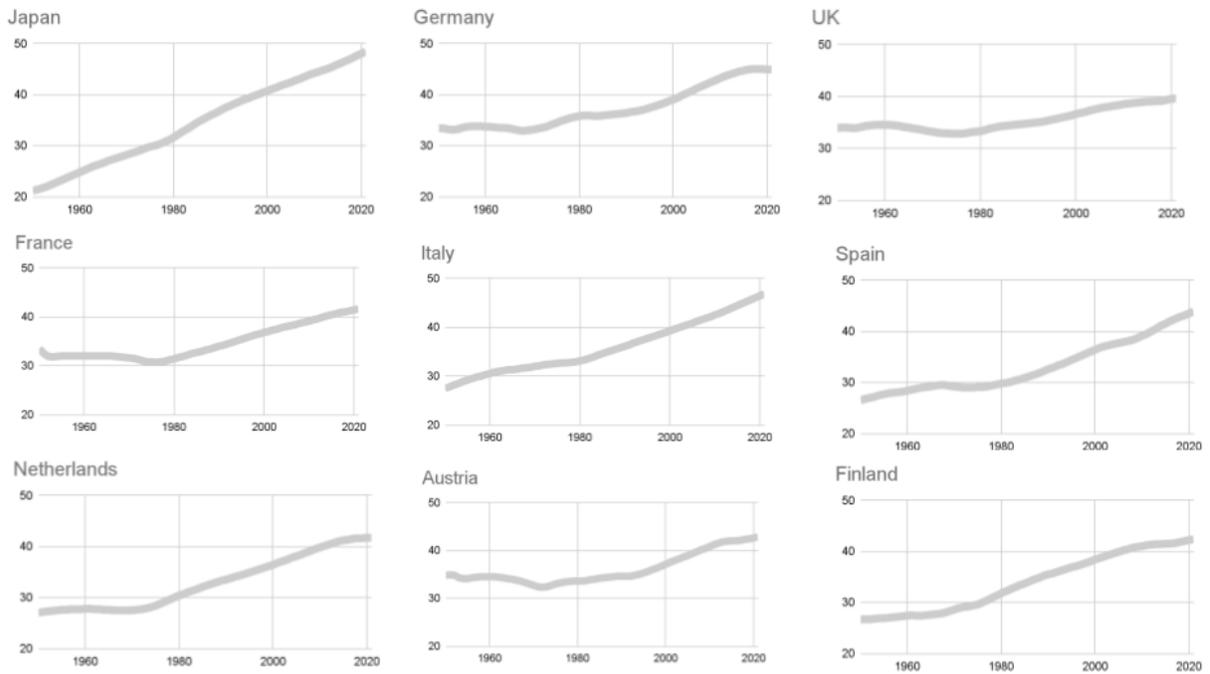

Fig. 2b. Median age of select industrialized countries 1950-2021

The following example illustrates our approach:

> Consider an economy where plain hats sell for $4 each, and carry a labor share of 0.2. Hats with customizable messages sell for $8 each. Consumers innovate on custom messages continuously, for sporting events, elections, alumni reunions, etc, leaving suppliers unable to automate the production process for custom hats beyond a labor share of 0.4. In short, consumers compete with suppliers to determine labor share.
>
> Younger consumers prefer custom hats. Older consumers find crafting custom messages onerous and prefer plain hats. Initially, both types of hats sell equally and the overall labor share is
>
> (0.5*0.2*4+0.5*0.4*8)/(0.5*4+0.5*8) = 0.33
>
> Now suppose the median age of the population increases, with relatively more older consumers, shifting the mix of hats to 60% plain hats. The overall labor share declines to
>
> (0.6*0.2*4+0.4*0.4*8)/(0.6*4+0.4*8) = 0.31
>
> In short, aging consumers demand fewer output variants, leading to a decline in labor share.

Our paper is organized is as follows:

1. We observe that human needs favor long tail distributions, demanding the same from goods and services that fulfill those needs, following Anderson (2004), Anderson (2006), Brynjolfsson et al (2007), Goel et al (2010) and Kausik (2023a). Based on this observation, we propose a theoretical model of economic activity on a long tail distribution, where innovation in demand for output variants on the long tail competes with innovation in supply automation at the head of the distribution. In other words, on the demand side, human consumers innovate in seeking new, niche variants of items that require human labor. While on the supply side, human producers innovate on automation to drive down the cost of mature, popular items.
2. We observe that human cognitive performance is stable in the working years, declining sharply thereafter, e.g., Salthouse (2010), Kobayashi et al, (2014), Murman (2015), and Eikelbloom et al. (2020). We invoke this observation to model innovation in supply and demand as a function of the median age of the population, then obtain an algebraic expression for labor share in a dynamic equilibrium.
3. We validate the model on historical data across industrialized economies via non-linear stochastic regression.

# Model of Economic Activity

We propose a theoretical macroeconomic model where economic activity follows a long tail distribution. Referring to Fig. 3, the horizontal axis is all available tasks in the economy ranked in decreasing order of economic output, where the rank fraction of a task is the ratio of the rank of the task over the total number of tasks. The vertical axis is the cumulative economic output share $P(r)$ of tasks with rank fraction between 0 and $r$. At any point in time, there exists rank fraction $a$ such that tasks in $[0, a]$ are performed by automation while tasks in $(a, 1]$ are performed by labor. We refer to $a$ as the automation fraction.

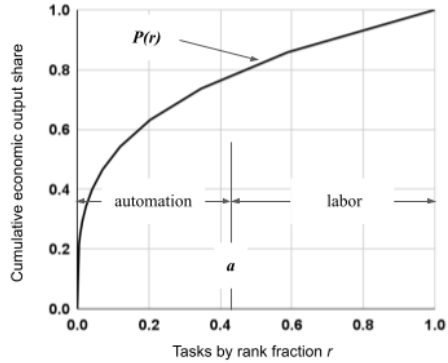

Fig. 3. Long tail model of economy

**Assumption 1:** The cumulative economic output share is given by $P(r) = r^n$ where $n$ is a constant model parameter. Automation's share of the output is $a^n$ and labor's share is $(1 - a^n)$.

**Assumption 2:** Tasks leftmost in the labor component are newly automated at a nominal rate $\sigma \geq 0$, in that the automation fraction $a$ nominally increases to $a + \sigma\tau$ after infinitesimal time interval $\tau$. See Fig. 4. We refer to $\sigma$ as the *supply innovation rate*.

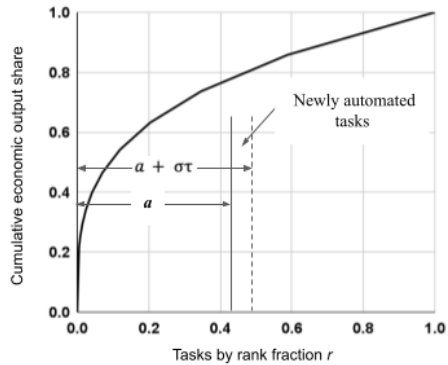

Fig. 4. Supply innovation: expanded automation of tasks

**Assumption 3:** As existing tasks become obsolete and new niche tasks are created, the rank fraction of tasks improves at a nominal rate $\delta > 0$, in that each point $r$ on the horizontal axis nominally shifts to $r(1 - \delta\tau)$ after infinitesimal time interval $\tau$. See Fig. 5. We refer to $\delta$ as the *demand innovation rate*.

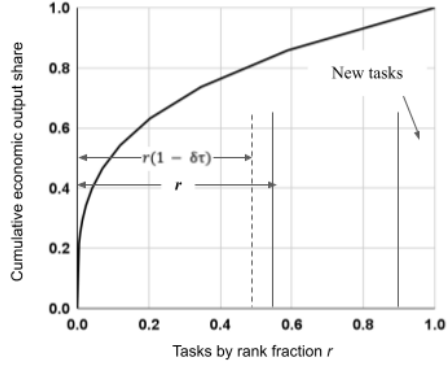

Fig. 5. Demand innovation: obsolescence and new niche tasks

**Assumption 4:** We assume that at time $t$,

$$\frac{\sigma}{\delta} = \left(\frac{\sigma_0}{\delta_0}\right)\frac{1}{1-k(\mu-\mu_0)}$$

where $\sigma_0$ and $\delta_0$ are the innovation rates at initial time $t_0$, $\mu$ and $\mu_0$ are the median age of the population at time $t$ and $t_0$ respectively, and $k \geq 0$ is a constant model parameter.

Assumption 4 is motivated by Fig. 6, which abstracts the literature on cognitive aging, e.g., Salthouse, (2010), Kobayashi et al, (2014), Murman (2015), and Eikelbloom et al. (2020). Per Fig. 6, cognitive performance, except vocabulary, is relatively stable during the working years but declines sharply thereafter. Since the supply innovation rate $\sigma$ pertains to the workforce and the demand innovation rate $\delta$ pertains to the entire population, Assumption 4 attenuates only the demand innovation rate $\delta$ by a factor linear in the median age of the population. We note that physical fitness also declines with age, but can impact the items consumed rather than the ability to demand variants of those items. In earlier work, Kausik (2023b), we assume ($\sigma/\delta$) is constant over the interval of interest, i.e., $k = 0$, leading to asymptotic transitions between static equilibria.

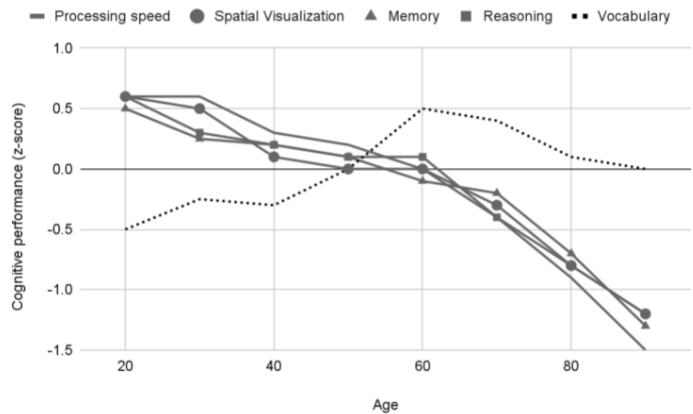

Fig. 6. Cognitive aging

The decline of human innovation with age is the subject of prior works, e.g Simonton (1988) and Jones (2010), who observe that innovation in science and technology, which is a combination of training and cognitive performance, peaks around age 40 and declines sharply thereafter. Likewise, in the context of internet search engines where query formulation corresponds to demand innovation and search results to output variants, prior works, e.g. Karanam and van Oostendorp (2016) and Kordovski et al. (2021), find that older adults have lower accuracy and lower rate of query formulation.

**Theorem 1:** In dynamic equilibrium, labor share is

$$1 - \left[\left(\frac{\sigma_0}{\delta_0}\right)\frac{1}{1-k(\mu-\mu_0)}\right]^n$$

**Proof:** Combining Assumptions 2 and 3, we get
$$\partial a/\partial t = \sigma - a\delta.$$
Rearranging, we get

$$\partial a/\partial t = -\delta\left(a - \frac{\sigma}{\delta}\right)$$

Setting $\partial a/\partial t = 0$ and noting $\delta > 0$, we get $a = \frac{\sigma}{\delta}$ at equilibrium, and hence by Assumption 4,

$$a = \left[\left(\frac{\sigma_0}{\delta_0}\right)\frac{1}{1-k(\mu-\mu_0)}\right]$$

Per Assumption 1, it follows that in dynamic equilibrium, the labor share is

$$1 - \left[\left(\frac{\sigma_0}{\delta_0}\right)\frac{1}{1-k(\mu-\mu_0)}\right]^n$$

# Validation

We now validate our model against historical economic data. Specifically, we use non-linear regression to show Theorem 1 maps median age to labor share for various countries ranging across population, industry, language, culture, geography and other factors. Our approach is similar to validating the hypothesis that the period of a pendulum of length $L$ is $T = 2\pi\sqrt{L/g}$, via non-linear regression between $T$, $L$ and $g$ across many pendulums and many planets with different gravitational constants $g$.

**Data Set 1:** We apply TensorFlow non-linear stochastic gradient descent regression to the time series of Figs. 1 and 2a to extract the parameters $(\sigma_0/\delta_0)$, $k$ and $n$, in the labor share formula of Theorem 1. Each run consists of 100 iterations of stochastic gradient descent and Mean Square Error loss, starting with the three parameters randomly initialized in the interval [0,1]. We average the three parameters across 20 runs to obtain the values shown below.

$$n = 7.86E - 01; \quad \frac{\sigma_0}{\delta_0} = 4.24E - 01; \quad k = 1.75E - 02$$

Fig. 7 plots the results. The thick gray line is the actual historical data 1950-2021. The solid black line is the model's fit across the same period. The Root Mean Squared error of the model in the period is ~2.4%.

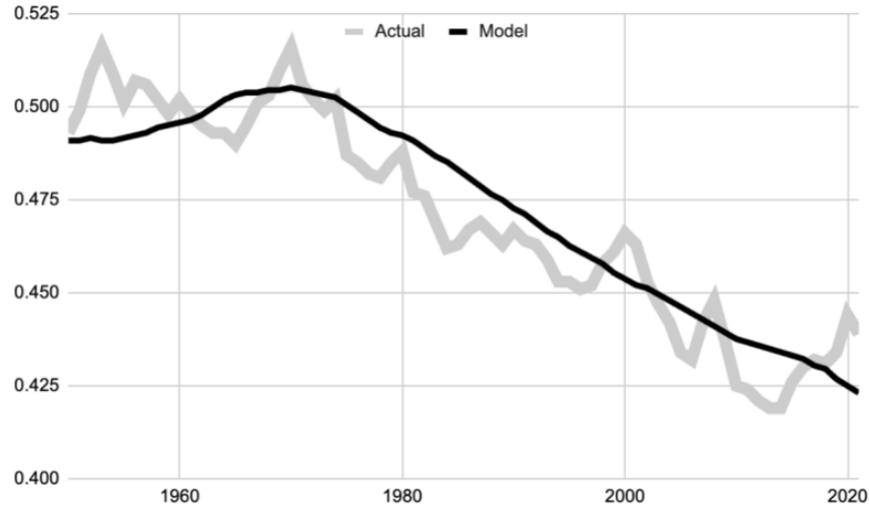

Fig. 7. Actual and model labor share of the United States

**Data Set 2:** We apply TensorFlow non-linear stochastic gradient descent regression to the time series of Fig. 2b and KLEMS 2013 economic data at http://www.euklems.net/. Each run consists of 100 iterations of stochastic gradient descent and Mean Square Error loss, starting with the three parameters randomly initialized in the interval [0,1]. We average the three parameters across 20 runs for each country to obtain the values shown in Table 1. Fig. 8 plots the results. For each country, the thick gray line is the actual historical data available between 1970 and 2012. The solid black line is the model's fit across the same period. The Root Mean Squared error of the model in the period is ~3% for each country.

| $n$ | $\sigma_0/\delta_0$ | $k$ | $n$ | $\sigma_0/\delta_0$ | $k$ | $n$ | $\sigma_0/\delta_0$ | $k$ |
|---|---|---|---|---|---|---|---|---|
| Japan | | | Germany | | | UK | | |
| 8.99E-01 | 2.63E-01 | 2.08E-02 | 9.04E-01 | 2.38E-01 | 2.10E-02 | 8.89e-1 | 2.51E-01 | 4.68E-03 |
| France | | | Italy | | | Spain | | |
| 9.62E-01 | 2.51E-01 | 4.10E-02 | 9.65E-01 | 2.41E-01 | 3.44E-02 | 8.85E-01 | 2.90E-01 | 1.33E-02 |
| Netherlands | | | Austria | | | Finland | | |
| 9.40E-01 | 2.54E-01 | 2.95E-02 | 9.65E-01 | 2.59E-01 | 3.70E-02 | 9.70E-01 | 2.28E-01 | 2.74E-02 |

Table 1. Model parameters for select industrialized countries

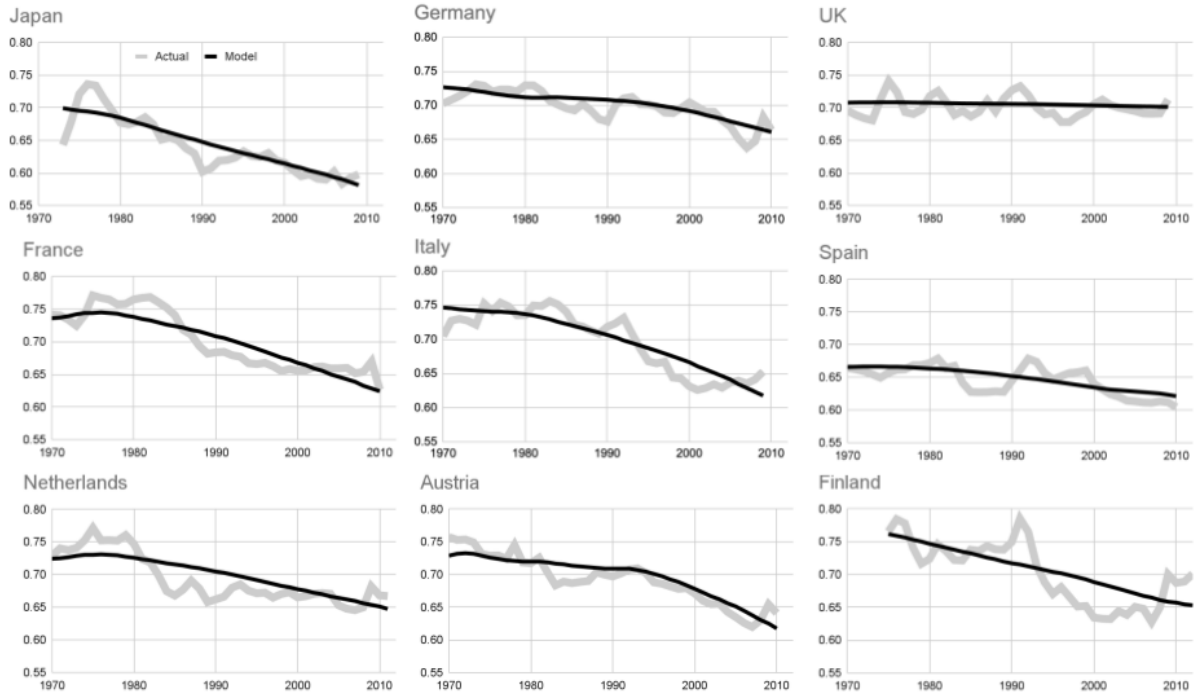

Fig. 8. Actual and model labor share of select industrialized countries

The foregoing shows Theorem 1's three parameter formula links labor share to median age across decades of historical data for many different countries with high accuracy, thereby validating the model.

We also seek corroboration of our model via the direct correlation between labor share and cognitive aging across industrialized countries. Since census measures of cognitive performance are not readily available, we define the decline in aggregate cognitive performance as the product of the increase in median age and the rate of decline in cognitive performance with age. Specifically,

$$Decline\ in\ aggregate\ cognitive\ performance\ (\%) =$$

$$increase\ in\ median\ age\ \times\ \frac{decline\ in\ average\ cognitive\ performance\ from\ ages\ 50s\ to\ 70s\ (\%)}{20\ years}$$

The above definition considers the decline in average cognitive performance from the age group 50-59 to the age group 70-79, skipping over the varying retirement rates in the age group 60-69.

Our data source for cognitive performance, https://g2aging.org, consolidates census measurements across countries, including US HRS, UK ELSA, EU SHARE, and Japan's JSTAR. For our interval of interest 1970-2012, 2006 and 2010 are the only years for which all four datasets include measurements on cognition. We average the 2006 and 2010 data for our analysis, and specifically, the average "Total Word Recall" score therein as our measure of average cognitive performance. Fig. 9 shows the results.

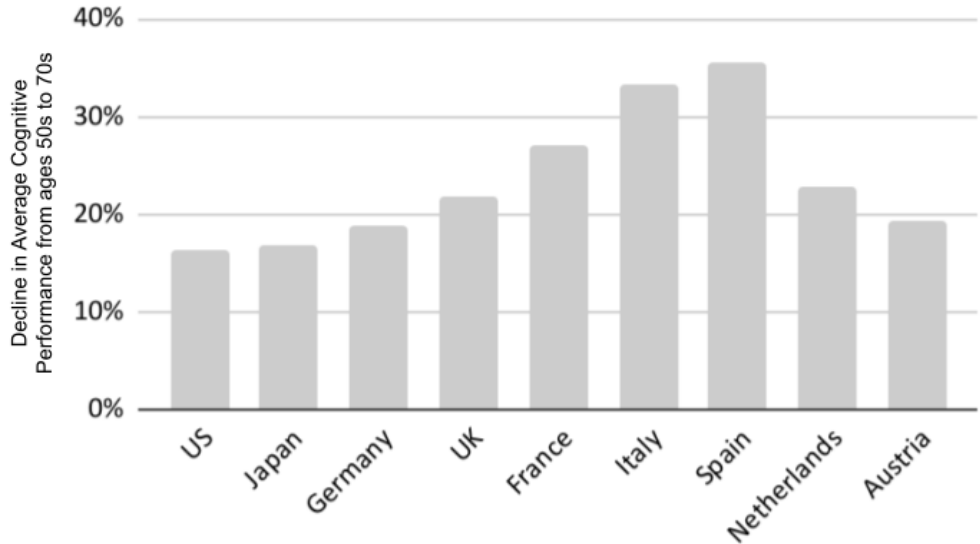

Fig. 9. Cognitive aging by country (average of 2006 & 2010).
*This graph uses data from the Gateway to Global Aging Data (g2aging.org). The Gateway to Global Aging Data is funded by the National Institute on Aging (R01 AG030153).*

For labor share, we use the same data as in Figs. 7 & 8, except that for consistency across countries we include a second data point for the US from the KLEMS 2013 dataset. Fig. 10 plots the decline in labor share vs the decline in aggregate cognitive performance across countries for which cognition data is available. The vertical axis shows the percentage decline in labor share as measured by linear regression on the available data for each country between 1970 and 2012. The data point labeled US-Fed represents the labor share data of Fig. 7 while all other data points represent labor share data from the KLEMS 2013 dataset of Fig. 8. The horizontal axis shows the decline in aggregate cognitive performance as calculated by combining Figs. 2a, 2b and 9 in the definition above.

During the years of interest, Spain trailed the others in industrialization with per capita GDP a fraction of the others. Omitting Spain as an outlier, for the remaining data points (a) the trend line is the linear regression through the origin, and (b) the correlation coefficient between the decline in labor share and the decline in aggregate cognitive performance is 0.65 and 0.69 with and without US-Fed respectively. For comparison, the correlation coefficient between the decline in labor share and just the increase in median age of Figs. 2a & 2b is 0.58 with or without US-Fed; and the correlation coefficient between the decline in labor share and just the decline in cognitive performance of Fig. 9 is 0.28 and 0.34 with and without US-Fed respectively.

An aging population can increase the demand for human labor from caregivers for older adults whose physical decline results in difficulty with the activities of daily living. However, data shows, e.g. https://g2aging.org, that cognitive aging begins at a much earlier age than physical decline, therefore affecting a greater share of the population. Against that background, we do not consider the impact of physical decline in our analysis.

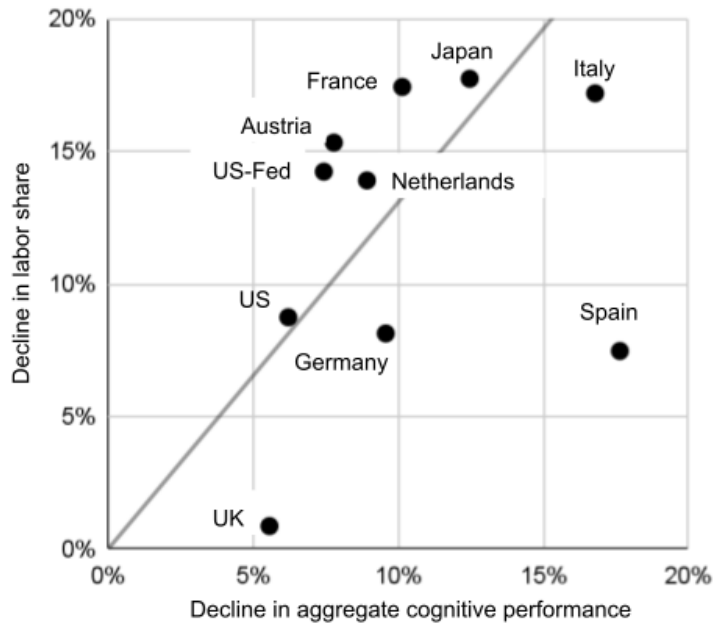

Fig. 10. Decline in labor share vs decline in aggregate cognitive performance (~1970~2012). *This graph uses data from the Gateway to Global Aging Data (g2aging.org). The Gateway to Global Aging Data is funded by the National Institute on Aging (R01 AG030153).*

# Summary

Labor share, the fraction of economic output accrued as wages, is inexplicably declining in industrialized countries. Whilst numerous prior works attempt to explain the decline via economic factors, our novel approach links the decline to biological factors. Specifically, we propose a theoretical macroeconomic model where labor share reflects a dynamic equilibrium between the workforce automating existing outputs, and consumers demanding new output variants that require human labor. Industrialization leads to an aging population, and while cognitive performance is stable in the working years it drops sharply thereafter. Consequently, the declining cognitive performance of aging consumers reduces the demand for new output variants, leading to a decline in labor share. Our model expresses labor share as an algebraic function of median age, and is validated with surprising accuracy on historical data across industrialized economies via non-linear stochastic regression.

Our analysis implies the cognitive performance of aging consumers is key to bolstering labor share. Other studies show that retirement can degrade cognitive performance, e.g., Bonsang et al. (2012), Adam et al. (2013), Nikolav and Shahadath Hossain (2023), Adam et al. (2007), and Rohwedder and Willis (2010),. Taken together, public policy has an economic interest in the cognitive health of aging retirees.